\documentclass[showpacs,prl,superscriptaddress,twocolumn,floatfix]{revtex4}
\usepackage{graphicx,float,amsmath,amssymb}

\def\be{\begin{equation}}
\def\ee{\end{equation}}

\def\bea{\begin{eqnarray}}
\def\eea{\end{eqnarray}}
\def\beann{\begin{eqnarray*}}
\def\eeann{\end{eqnarray*}}
\def\I{{\rm i}}                  
\def\D{{\rm d}}                  
\newcommand{\mean}[1]{\left\langle #1 \right\rangle} 
\newcommand{\EXP}[1]{{\mbox{\large e}}^{#1}}         

\newcommand{\drond}[2]{\frac{\partial #1}{\partial #2}} 

\begin{document}

\title{Direct measurement of the phase coherence length in a GaAs/GaAlAs
        square network}

\author{M. Ferrier}
\affiliation{Laboratoire de Physique des Solides, Associ\'e au CNRS,
              Universit\'e Paris-Sud, 91405 Orsay, France.}
\author{L. Angers}
\affiliation{Laboratoire de Physique des Solides, Associ\'e au CNRS,
              Universit\'e Paris-Sud, 91405 Orsay, France.}
\author{A.C.H. Rowe}
\affiliation{Laboratoire de Physique des Solides, Associ\'e au CNRS,
              Universit\'e Paris-Sud, 91405 Orsay, France.}              
\author{S. Gu\'eron}
\affiliation{Laboratoire de Physique des Solides, Associ\'e au CNRS,
              Universit\'e Paris-Sud, 91405 Orsay, France.}
\author{H. Bouchiat}
\affiliation{Laboratoire de Physique des Solides, Associ\'e au CNRS,
              Universit\'e Paris-Sud, 91405 Orsay, France.}
\author{C. Texier}
\affiliation{Laboratoire de Physique Th\'eorique et Mod\`eles Statistiques,
              Associ\'e au CNRS, Universit\'e Paris-Sud.}
\affiliation{Laboratoire de Physique des Solides, Associ\'e au CNRS,
              Universit\'e Paris-Sud, 91405 Orsay, France.}
\author{G. Montambaux}
\affiliation{Laboratoire de Physique des Solides, Associ\'e au CNRS,
              Universit\'e Paris-Sud, 91405 Orsay, France.}
\author{D. Mailly}
\affiliation{Laboratoire de Photonique et Nanostructures, CNRS,
              Route de Nozay, 91460 Marcoussis, France.}


\date{December 10, 2004}

\begin{abstract}
The low temperature magnetoconductance of a large array of quantum coherent
loops  exhibits Altshuler-Aronov-Spivak  oscillations  which periodicity
corresponds to 1/2 flux quantum per loop.
We show that the measurement of the harmonics content
provides an accurate way to determine the electron phase coherence length
$L_{\varphi}$   in units of the lattice  length with no  adjustable
parameters.
We use this method to determine $L_{\varphi}$ in a square network realised 
from a 2D
electron gas (2DEG) in a GaAS/GaAlAs heterojunction, with only a few 
conducting channels. The temperature
dependence follows a power law $T^{-1/3}$ from 1.3 K to 25 mK with no
saturation, as expected for 1D diffusive electronic motion and
electron-electron scattering  as the main  decoherence mechanism.
\end{abstract}




\maketitle

The characteristic scale on which quantum interference can occur
in a conductor, the phase coherence length $L_{\varphi}$, is the
key parameter of quantum  transport.
In particular, the dependence of $L_{\varphi}$ on temperature can
discriminate between the  various scattering mechanisms which limit phase
coherence: electron-electron (e-e), electron-phonon or electron-magnetic
impurity interactions. Interference on the scale of $L_{\varphi}$ gives rise
to  two  different types of contributions  to the conductance  in a transport
experiment.
Some are sample specific and depend on the particular disorder
configuration. These are conductance fluctuations (magnetofingerprints) and
the $\phi_0$ periodic Aharonov-Bohm (AB) oscillations ($\phi_0=h/e$ is the
flux quantum).
Their amplitudes are governed both by $L_{\varphi}$ and the thermal length
$L_{T}$, in general smaller than $L_{\varphi}$. This makes an
accurate determination of $L_{\varphi}$ difficult
\cite{AltAro85,EchGerBozBogNil93,PieGouAntPotEstBir03}. 
The second type of contribution, called the weak localisation (WL)
correction, is obtained after ensemble averaging of quantum interferences on
many configurations of disorder. It originates from interferences between
time reversed electronic trajectories,
which are the only ones surviving the disorder average. It is also observed
in samples of size $L\gg(L_{\varphi},L_{T})$ and only depends on $L_{\varphi}$
since it involves trajectories at the same energy.
Manifestations of WL are the magnetoconductance (MC) of large connex samples
\cite{AltAro85,EchGerBozBogNil93,PieGouAntPotEstBir03} and the
Altshuler-Aronov-Spivak (AAS) $\phi_0/2$ periodic oscillations resulting
from  the ensemble average of AB oscillations in a long cylinder or large
arrays of connected phase coherent rings
\cite{AroSha87,PanChaRamGan84,DolLicBis86}.
The WL provides thus in general a much more direct measurement of
$L_{\varphi}$ than sample specific corrections.

\begin{figure}
\includegraphics[clip=true,width=7.5cm]{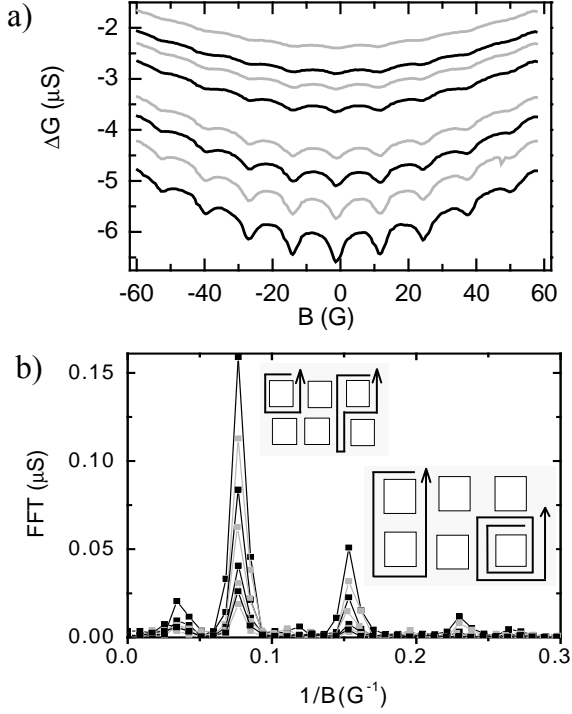}
\caption{a) Conductance versus magnetic field between
$25\:$mK to $220\:$mK.
b) Fourier transform of the MC (after substraction of the envelope) for different temperatures.
Left (right) inset: some orbits contributing to the first (second) harmonic.
\label{dGdeT}}
\end{figure}

The analysis of the MC in 1D diffusive metallic wires (with
transverse dimensions smaller than $L_{\varphi}$) has led
to accurate determinations of $L_{\varphi}$. It was found that the dominant
phase breaking mechanism at very low temperature,  in the absence of magnetic
impurities, is due to e-e scattering and is well described by
the Altshuler-Aronov-Khmelnitskii (AAK) theory  \cite{AltAroKhm82,AltAro85}
yielding $L_{\varphi}\propto T^{-{1}/{3}}$ with no saturation down to $40\:$mK
\cite{EchGerBozBogNil93,PieGouAntPotEstBir03}.
Such a  remarkable agreement  between theory and experiment has not been
established for semiconducting wires, where most WL experiments have been
performed only above $0.2\:$K  or with insufficient ensemble averaging
\cite{ReuBouMai95,ThoPepAhmAndDav86, vantout88, gershen95}. It is however
essential  to check the validity of the AAK theory for these systems which
correspond to  radically different physical parameters:  fewer  conducting
channels and  larger screening lengths. In this Letter  we present
MC data down to 25 mK of  networks fabricated from a
GaAs/GaAlAs  2DEG, which contain $10^6$ square loops in the
diffusive transport regime, and  determine $L_{\varphi}$   {\it without
adjustable parameters} from the analysis of the AAS oscillations
(Fig.~\ref{dGdeT}).  Following 
\cite{DouRam85,PasMon99}, we explain how to calculate
the harmonics content of these oscillations and show  that it depends only on
$L/L_{\varphi}$where $L$ is the circumference of the elementary loop.  It is then possible to determine 
$L_{\varphi}$  and its temperature dependence exclusively from geometrical parameters of the network.
This new method of determinating $L_{\varphi}$ is especially  interesting in these 2DEG wires for which basic transport parameters such as the electron density  and wire width ($W$)  are not straightforwardly determined, unlike metals. 

Moreover once $L_{\varphi}$  is determined, we deduce from the
analysis of the high field positive MC  the elastic mean free path
($l_{e}$), W, and make a detailed comparison with
theoretical predictions of the AAK theory on dephasing by e-e
interactions. We find a very good quantitative  agreement in the regime, never explored
before, of very few conducting channels.

In the  weakly disordered diffusive regime ($k_F l_e \gg1$),
the WL correction is directly related  to the Cooperon, which can be
computed from the time integrated return probability
$P_c(\vec r,\vec r)$ for a diffusive electron. In a cylinder or an array of
connected loops the contribution  to the Cooperon of trajectories enlacing at
least one loop oscillates with a flux periodicity of ${\phi_{0}}/{2}$
giving rise to the AAS oscillations. A systematic way of calculating WL in a
mesoscopic network of diffusive wires was derived in
\cite{DouRam85}. 
More recently \cite{PasMon99}, a relation was found
between the WL correction  and the spectral determinant
$S(\gamma)=\det(\gamma-\Delta)$ of the Laplace operator $\Delta$ defined on
the network. If we write
$\Delta\tilde{\sigma}=\mean{\Delta\sigma}h/e^2$, then
\be\label{deltasigma}
\Delta\tilde{\sigma}=
-4\int\frac{\D\vec r}{{\rm Vol}}\,P_c(\vec r,\vec r)
=
\frac{-4}{{\rm Vol}}\frac{\partial}{\partial\gamma} \ln S(\gamma)
\:,\ee
where $\gamma={1}/{L_{\varphi}^{2}}$. 
Eq.(\ref{deltasigma}) assumes an exponential relaxation of phase coherence.
This approach, which is meaningful 
only for regular networks, is particularly efficient because $S(\gamma)$ can
be computed systematically for any given network in terms of the determinant
of a finite size matrix encoding the network's characteristics
(topology, length of the wires, magnetic flux).
It can also be shown that the WL can be expressed, in the small $L_{\varphi}$
limit, as a trace expansion over periodic orbits, denoted ${\cal C}$,
$$
\hspace{-0.15cm}
\drond{\ln S(\gamma)}{\gamma}=\frac{{1}}{2\sqrt\gamma}
\left[{\cal L} +\frac{V-B}{\sqrt\gamma}
+\sum_{{\cal C}} l(\tilde{\cal C})
\alpha({\cal C})\EXP{-\sqrt{\gamma}\,l({\cal C})+\I\theta({\cal C})}\right]
$$
where $V$ ($B$), is the total number of vertices (bonds).
$\tilde{\cal C}$ is the primitive orbit related to ${\cal C}$.
${\rm Vol}={\cal L}$ is the total length.
We explain briefly this formula, demonstrated in \cite{Rot83} and discussed in
detail in \cite{AkkComDesMonTex00}.
Each orbit contributes to the MC with a phase factor which
depends on the enclosed flux: $\theta({\cal C})= 4\pi\Phi({\cal C})/{\phi_0}$.
It is  also characterised  by its length $l({\cal C})$ and by a geometrical
weight $\alpha({\cal C})$. In the case of a square lattice
of periodicity  $a$, the periodically oscillating  conductance can be 
decomposed
in Fourier space as a sum of harmonics of the fundamental periodicity
corresponding to $\phi_0/2$ per elementary cell.
The first terms of this expansion in $x=\EXP{-2a/L_\varphi}$ read:
\bea\label{sigmaforsquare}
\Delta\tilde{\sigma} =
- 
\frac{L_\varphi}{W}
\bigg[ \hspace{-.4cm}&&
   2 - \frac{L_\varphi}{a}+
   x+\cdots  
     +\frac{x^{2}}{2}\cos\theta\:(1-\frac32\:x+\cdots)
   \nonumber\\
   &&+\ \frac{3\:x^{3}}{8}\cos2\theta\: (1-
\frac{19}{12}\:x+\cdots) \nonumber\\
   &&+\ \frac{3\:x^{4}}{8}\cos3\theta\:(1-
\frac{15}{8}\:x+\cdots)\bigg]
.\eea
Here $\theta = 4\pi{\phi}/{\phi_{0}}$, and $\phi$ is the flux per elementary
cell.
The amplitude of the $n$th harmonics is evaluated by counting the paths
enclosing $n$ fluxes $\phi$.
The counting is rapidly cumbersome (156 orbits are
involved in the last term),
but the crucial point is that the coefficient of each term depends only on the
lattice geometry.

\begin{figure}
\includegraphics[clip=true,width=8 cm]{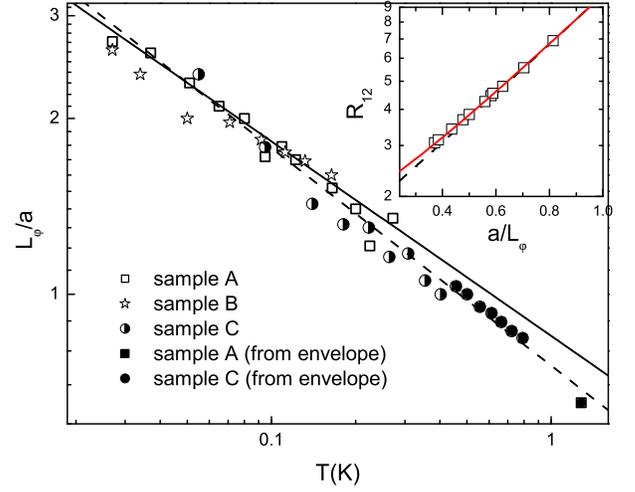}
\caption{inset: relation between ${L_{\varphi}}/{a}$ and the ratio $R_{12}$ of
the two first harmonics on a  semi-log scale.
The continuous line comes from the numerical calculation of $S(\gamma)$.
The dashed line is deduced
from  the expansion (\ref{sigmaforsquare}). The open circles are the
experimental values of $R_{12}$ from which $L_{\varphi}$ is determined. \\
Main panel: $L_{\varphi}$ versus temperature on a log-log scale obtained from $R_{12}$ for samples A,B and C. The
fit (dashed line) yields the power law $L_{\varphi} \propto T^{-0.36}$. The dark circles and squares  are obtained from the fit  of the envelope.
The continuous line is $L_\varphi$ from the AAK theory (for sample A).
\label{LphideT}}
\end{figure}

More generally, the WL correction can be obtained for all values of
${L_{\varphi}}/{a}$ from the numerical computation of the
determinant in Eq.~(\ref{deltasigma}). The numerical FFT of
the computed MC yields the ratio $R_{12}$ of the 2 first
harmonics as a function of $L_{\varphi}/a$ (Fig.~\ref{LphideT}a). It
appears that the small orbit expansion (\ref{sigmaforsquare}) is a good
approximation up to ${L_{\varphi}}/{a} = 2$. In any case the ratio of two
harmonics  is completely determined by ${L_{\varphi}}/{a}$ and provides
a  method for a direct evaluation of $L_{\varphi}$ without any
adjustable parameter.
The square lattice is particularly appropriate for such a determination of
$L_{\varphi}$ due to its large harmonics content: the second harmonic is
dominated by orbits of length $6a$ instead of $8a$ for a statistical ensemble
of single rings or a necklace of identical rings, for example.

We now use this method to determine the phase coherence length
of square networks etched in a 2DEG of a
GaAs/AlGaAs heterostructure. The networks consist of $10^6$ square loops of
side $a=1\:\mu$m and nominal width $W_{0}=0.5\:\mu$m and cover a total area of
$1\:$mm$^2$. A gold gate deposited $100\:$nm above the 2DEG offers
the possibility to change the number of electrons in the
network.
Measurements were done on three networks (A,B with gate, C without ), giving the same
results. Except when specified, figures show the data for sample A.
We have measured the MC up to $4.5\:$T between $25\:$mK and $1.3\:$K, using
a standard lock-in technique  (ac current of $1\:$nA at $30\:$Hz). The samples were in general strongly depleted at low temperature because of
the etching. The intrinsic electron density of the 2DEG,
$n_e=4.4\times10^{15}\:$m$^{-2}$, was recovered after
illuminating the samples during several minutes at $4.2\:$K.
This density   was  determined  from Shubnikov-de Haas
oscillations  visible above $1\:$T. 
Because of depletion after etching of the 2DEG, it is difficult to
estimate the  real width of the wires (W) and $l_e$. 

\begin{figure}
\includegraphics[clip=true,width=8cm]{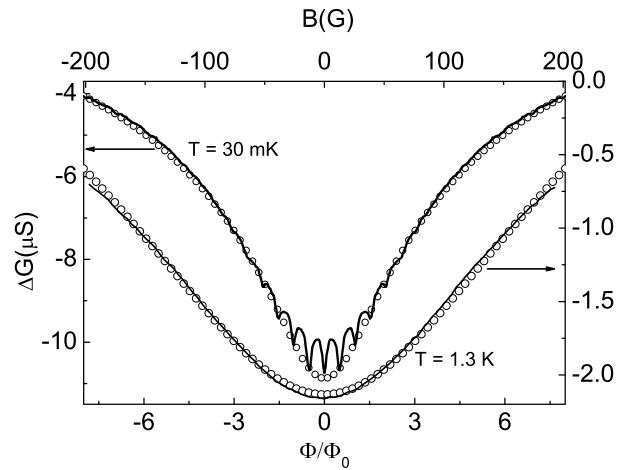}
\caption{
High field MC, the continuous lines are the experimental data, the dots
  are the fits with eq.~(\ref{env}).
Parameters of the fits are~: $l_{e}$ and $W$ at $30\:$mK, 
and $L_{\varphi}$ and $l_{e}$ at $1.3\:$K.
\label{RfortH}}
\end{figure}

At low magnetic field (Fig.~\ref{dGdeT}a),
the MC exhibit large AAS oscillations  with a period  $12.6\:$G
corresponding to a flux ${\phi_{0}}/{2}$ in a square cell of area
$a^2$.  The oscillations are clearly not purely sinusoidal.  At the
lowest temperature, $25\:$mK, three harmonics are visible in the Fourier
spectrum of the MC (Fig.~\ref{dGdeT}b).
Moreover, as shown in Fig.~\ref{RfortH} which  represents the MC
for a wider range of field, the oscillations disappear above
$60\:$G but the WL magnetoconductance due to the penetration of the 
field  through the finite width of the wires constituting the network is
still clearly visible.
At high temperature, above $400\:$mK, the AAS oscillations disappear even at
low field. Only the positive MC remains with a smaller amplitude. In sample B,
the same experiments for different gate voltages were also performed.

We first concentrate on the  AAS oscillations (Fig.~\ref{dGdeT}a).
The Fourier spectrum of  the MC exhibits a series of peaks
corresponding to successive harmonics of the $\phi_0/2$
periodicity. The finite width of the peaks (Fig.~\ref{dGdeT}b) is due to the
penetration of the magnetic field in the wires which damps the AAS
oscillations at high field.
It can be shown that this broadening does not affect the integral of the peak.
A first rough analysis shows that the ratio $R_{12}$ of integrated peaks of
the two first harmonics behaves like $R_{12}\sim\exp T^{1/3}$.
We now use the theory described above to
quantitatively determine $L_{\varphi}$ via the 
relation between $L_{\varphi}/a $ and $R_{12}$. We deduce its temperature
dependence  between $25\:$mK and $250\:$mK as shown in
Fig.~\ref{LphideT}. We find that
$L_{\varphi}$ follows a power law $T^{-\eta}$, where $\eta= 0.36 \pm 0.05$.
The coherence length reaches almost $3\:\mu$m at $25\:$mK and there is no
sign of saturation.

Once $L_{\varphi}$ is determined, the sample parameters
($W$, $l_{e}$) can be deduced from the WL envelope.
The magnetic field appears as an additional
effective phase breaking rate  for the time reversed trajectories
responsible for the WL leading to an effective $L_{\varphi}$ given by
\cite{AltAro81}:
\be\label{Lphi_eff}
\frac{1}{L_\varphi({\cal\phi})^2} = \frac{1}{L_\varphi(0)^2}
\left[ 1+\frac{1}{3}\left(2\pi\frac{\phi}{\phi_0}
\frac{W_{\rm eff}L_\varphi(0)}{a^2}\right)^2 \right]
\ee
where $W_{\rm eff}= W \sqrt{(3W)/(C_1l_e)}$ is  a
renormalized width
which appears in the WL  correction for a semi ballistic wire
($l_e\gg W$) due to  the phenomenon of flux cancellation. 
The coefficient $C_1$ depends on the specific boundary conditions.
The samples under consideration are close to the case of specular boundary
scattering \cite{ReuBouMai95} for which $C_1=9.5$.
The MC envelope, given by
$\left\langle \Delta\sigma(\phi=0,L_\varphi(\phi)) \right\rangle$, can 
be analytically
computed for the square lattice geometry and is given by:
\be\label{env}
\hspace{-0.3cm}
 \Delta\tilde{\sigma} = -
\frac{L_\varphi}{W} \left[ \coth {a  \over
L_\varphi} - {L_\varphi \over a} + {2 \over \pi} \tanh \frac{a}{L_\varphi}\,
K\!\left({1 \over \cosh\frac{a}{L_\varphi}}\right)\right]
\ee
where $K(x)$ is a complete elliptic integral. This expression is used in a
2-parameter ($W$, $\sigma_{\rm D}$) fit of $\Delta\sigma / \sigma_{\rm D} =
\Delta G / G_{\rm D}$ where $\sigma_{\rm D}$ and $G_{\rm D}$ are the Drude
conductivity and conductance. Since $k_{F}$ is determined independently from
Shubnikov-de Haas measurements, $\sigma_{\rm D} = \frac{e^2}{h}k_{F}l_{e}$
determines $l_{e}$. The above expression for $W_{\rm{eff}}$ can then be used
to find $W$. For sample A/B/C, $W = 170/270/230\:$nm and 
$l_{e} = 220/250/360\:$nm, independent of temperature as expected. This also
shows that the networks are indeed in the diffusive regime. In sample A the
number of transverse channels per wire is $M=k_FW/\pi=9$, and the number of
effective conducting channels on the scale of $a$ is only 
$M_{\rm eff}= Ml_e/a\sim2$. Results for samples B and C are similar.

At higher temperature where no AAS oscillations are visible, we can
nevertheless deduce $L_{\varphi}$ and $l_{e}$ by fitting the MC, knowing the temperature independent values
of $W$ and $G_{\rm D}$ (Fig.~\ref{RfortH}). Thus a quantitative comparison of
$L_{\varphi}$ with the theoretical prediction of
AAK \cite{AltAroKhm82,aleiner99}
$L_\varphi=\sqrt2(\frac{D^2m^*W}{\pi k_BT})^{1/3}$ written for a 2DEG wire is possible.
This theory applies to diffusive metallic wires with a large number of
conducting channels in a limit where e-e interactions are treated
perturbatively.
We find a very good agreement (see Fig.~\ref{LphideT}b) which
is remarquable for two reasons:
({\it i}) we are confronted here with a small number of conducting channels
where strong interaction effects could be expected,
({\it ii}) the result of AAK was not extended to network geometry.
Recently it was predicted in \cite{LudMir03} that $L_\varphi$ extracted from the
AB or AAS oscillations in a single ring of perimeter $L$ should behave
like $L_\varphi\propto(LT)^{-1/2}$ corresponding to
$R_{12}\sim\exp L^{3/2}T^{1/2}$. This behaviour is not observed in our 
experiment.

For sample B
gate voltages between $-0.3\:$V and $0.3\:$V changed the
resistance from $30\:$k$\Omega$ to $400\:$k$\Omega$. A good filtering of the
gate voltage line is needed to avoid saturation of $L_{\varphi}$. Within our experimental accuracy, we find that $L_{\varphi}$
is not changed  and still varies as  $T^{-1/3}$. We
estimated for each gate voltage $W$ and $l_e$. When the
gate voltage varies between
$0.15\:$V and $-0.15\:$V, $W$ is unchanged  but
$l_e$  decreases  by a factor $1.5$ and $n_e$ by $30\%$; $G_{\rm D}$ decreases by a
factor $5$. This shows that the effect of the gate is   mainly to disconnect
bonds of the network.

\begin{figure}
\includegraphics[clip=true,width=8cm]{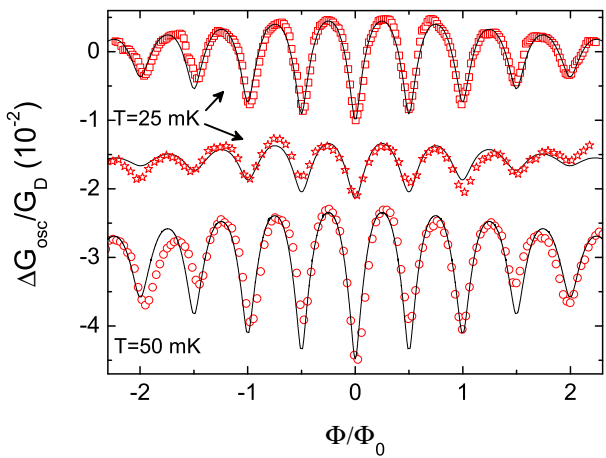}
\caption{
Comparison between experiment (symbols) and theory (continuous line) for the
oscillating part of the conductance for sample A ($\square$)($G_D =
3.7\times 10^{-5}$S) ,
sample B ($\star$)($G_D = 5\times 10^{-6}$S) with $V_g=-0.15\:$V corresponding to $W=240\:$nm and $l_e = 170\:$nm and sample C ($\bigcirc$)($G_D = 2.6\times 10^{-5}$) (shifted down for visibility). 
The only adjustable parameter is the amplitude of the oscillations.
\label{fitcomp}
}
\end{figure}

As a consistency check we have computed numerically the  oscillating part of
the MC with formula (\ref{deltasigma}) and (\ref{Lphi_eff})
using the value of $W_{\rm eff}$ determined above from the WL
envelope of the MC curves. We find that this value also precisely describes
the damping  of the AAS  oscillations, if the oscillations amplitude
is multiplied by a factor ranging from 1.6 to 2 depending on the
gate voltage. This can be explained by the existence of broken bonds in the
network which influences envelope and oscillations differently \cite{CT}. We obtain
a very good agreement between theory and experiments (Fig.~\ref{fitcomp}).

In conclusion we have shown that magnetoconductance experiments in GaAs/GaAlAs
networks can be described very accurately by the diagrammatic theory of
quantum transport in diffusive networks. It is remarkable that this
agreement is achieved in a limit where the  dimensionless conductance on the
scale of the  period of the network, $Ml_e/a$, is of the order 1, and
down to temperatures corresponding to $L_{\varphi}\sim L_{T}$ {\it i.e.} 
close to the limit of validity of AAK theory.
In contrast, metallic wires  deep in the diffusive regime have a
number of  conducting channels of order 
1000 and $L_{\varphi}\gg L_{T}$ is always fulfilled. 
Moreover we extracted from the AAS oscillations the temperature dependance of
the phase coherence length $L_{\varphi}\propto T^{-1/3}$ that agrees with
AAK theory down to $25\:$mK.

We thank B.~Etienne for the heterojunctions, and R.~Deblock
and B.~Reulet for fruitful discussions.



\begin{thebibliography}{100}
\bibitem{CT}
The number of broken bonds is estimated from the resistance of the network
to be of order 10\%. This only weakly affects the analysis of the MC.
\end{thebibliography}
\end{document}